\documentclass[11pt]{article}
\usepackage{graphicx}
\graphicspath{{C:/Users/xinwei/Documents/6EPMapping/gunterPaper/}}

\columnwidth\textwidth

\def\fii{\varphi}
\def\al{\alpha}

\def\ro{\varrho}
\def\si{\sigma}

\def\d{\partial}
\def\=d{\,{\buildrel\rm def\over =}\,}

\def\sqr#1#2{{\vcenter{\vbox{\hrule height.#2pt\hbox{\vrule width.
#2pt height#1pt \kern#1pt \vrule width.#2pt}\hrule height.#2pt}}}}

\def\B{\Bigl}

\begin{document}

\title{The hyperbolic heat transfer equation and the ablation problem: Theory and experiment}

\author{G\"unter Scharf\footnote{e-mail: scharf@physik.uzh.ch}
\\ Physics Institute, University of Z\"urich
\\Lam Dang\footnote{e-mail: lam.dang@gmail.com}
\\HerzGef\"assZentrum, Klinik im Park
\\8022 Z\"urich}

\date{ }

\maketitle\vskip 3cm

\vskip 3cm

We study the ablation problem for the hyperbolic heat equation in an axisymmetric  geometry which can be conveniently realized in the lab.
We determine an analytic solution which shows the approach to steady state. The thermal relaxation time $\tau$ is best obtained from the small time
behavior. The measurements give a surprisingly large $\tau$ of about 7 minutes for 0.5 \% NaCl in water. This shows that the hyperbolic equation must certainly be used instead of the parabolic heat equation in the ablation problem of electrocardiology.

\newpage
\section{Introduction}

The hyperbolic heat transfer equation
$${\d^2 T\over\d t^2}+{1\over\tau}{\d T\over\d t}-{\kappa\over\tau\ro c}\triangle T={1\over\ro c}{\d Q\over\d t}+{Q\over\tau\ro c}.\eqno(1.1)$$
has been applied to the ablation problem in previous studies ([1] and references given there). In (1.1) $T$ is temperature, $t$ is time, $\triangle$ is the Laplace operator and $Q$ describes the heat generation. The constant $\kappa$ is the thermal conductivity, $\ro$ the mass density and $c$ the specific heat of the medium. 
 However the value of the thermal relaxation time $\tau$ is badly known and consequently, it is unclear whether it is necessary to use (1) instead of the simpler parabolic heat equation. In [2] a value $\tau =16$ seconds has been measured in ``processed meat'' by methods using thermal conduction only, without heat generation. This value then has been used also in simulations of the ablation problem [1]. But if Joule's heat is generated by an electric current the physics is completely different from pure heat conduction in [2]. The electric energy is first transferred to kinetic energy of moving charged ions (Na+ and Cl- in blood or water). Then it is dissipated to the water molecules where it is measured by the thermometers. This total relaxation process is much slower. In fact, we shall see that $\tau$ is now measured in minutes instead of seconds as in [2]. This means that we must certainly use equation (1.1) for the ablation problem in electrocardiology.

\section{The axisymmetric ablation problem}

Let us consider a cylindrical electrode of radius $r_0$ and length $L$ with potential $V_0$ in a medium with electrical conductivity $\si$. A second dispersive cylindrical electrode  with radius $r_1$ ($r_1\gg r_0$) is grounded with potential $V=0$. The potential $V(r)$ in the medium is given by the simple solution of the two-dimensional Laplace's equation
$$V(r)=C_0+C_1\log r,\eqno(2.1)$$
if boundary effects at the end of the electrodes are neglected. Using the boundary conditions
$$V(r_0)=V_0,\quad V(r_1)=0\eqno(2.1)$$
we have
$$V(r)=V_0{\log r-\log r_1\over\log r_0-\log r_1}.\eqno(2.3)$$
The corresponding electric field strength is
$$E_r=-{\d V\over\d r}=-{V_0\over \log r_0-\log r_1}{1\over r}\eqno(2.4)$$
and the heat generation is equal to
$$Q=\si E_r^2={\beta\over r^2}\quad {\rm with}\quad \beta=\si{V_0^2\over(\log r_0-\log r_1)^2}={P\over 2\pi L\log^2(r_0/r_1)}\eqno(2.5)$$
where $P$ is the electric power in watts. With this heating we want to calculate the transient temperature $T(t,r)$ as solution of equation (1.1) in two dimensions.

We write the solution as
$$T(t,r)=T_1(r)+T_2(t,r),\eqno(2.6)$$
where $T_1(r)$ is the steady state solution satisfying
$${d\over dr}\B(r{dT_1\over dr}\B)=-{\beta\over \kappa r}.\eqno(2.7)$$
This is easily integrated
$$T_1(r)=-{b\over 2}(\log r)^2+C_2\log r+C_3\eqno(2.8)$$
where $b=\beta/\kappa$. The integration constants are fixed by the assumption of no heat flux at the electrode
$${dT_1\over dr}\B\vert_{r_0}=0,\eqno(2.9)$$
which is reasonable for a thin electrode, and by the second boundary condition
$$T_1(r_1)=T_0.\eqno(2.10)$$
 This gives the following steady state solution
$$T_1(r)=T_0+b\log{r_0\over r_1}\log{r\over r_1}-{b\over 2r^2}\log^2{r\over r_1}.\eqno(2.11)$$
The maximal temperature is found at the electrode, of course
$$T_1(r_0)=T_0+{b\over 2\log^2{r_1\over r_0}}.\eqno(2.12)$$

The remaining homogeneous equation for $T_2$ is solved by separation of variables
$$T_2(t,r)=T_3(t)T_4(r).\eqno(2.13)$$
Then we have
$${1\over a}{\dot T_3\over T_3}+a\tau{\ddot T_3\over T_3}={T_4''\over T_4}+{1\over r}{T_4'\over T_4}={\rm const.}=-\,k^2\eqno(2.14)$$
where the dot means time derivative and the prime radial derivative and
$$a={\kappa\over\ro c}.\eqno(2.15)$$
This yields
$$a\tau\ddot T_3+{1\over a}\dot T_3=-k^2T_3\eqno(2.16)$$
and
$$r^2T_4''+rT_4'=-k^2r^2T_4=0.\eqno(2.17)$$
This last equation is a special case of Bessel's equation [3]. We transform it to a self-adjoint form
$$y(r)=\sqrt{r}T_4(r)\eqno(2.18)$$
so that
$$-y''-{y\over4r^2}=k^2y.\eqno(2.19)$$
A fundamental system of solutions is given by
$$y(r)=\sqrt{r}(D_1J_0(kr)+D_2Y_0(kr))\eqno(2.20)$$
where $J_0$ and $Y_0$ are the Bessel functions of first and second kind.

The integration constants in (2.20) are determined by the two boundary conditions
$$T_4(t,r_1)=0,\quad {\d T_4\over\d r}\B\vert_{r_0}=0\eqno(2.21)$$
which lead to the self-adjoint boundary conditions
$$y(r_1)=0,\quad r_0y'(r_0)-{1\over 2}y(r_0)=0.\eqno(2.22)$$
The first one yields
$$D_2(k)=-{J_0(kr_1)\over Y_0(kr_1)}D_1(k)$$
and the second one gives
$$D_1(k)J'_0(kr_0)+D_2(k)Y'_0(kr_0)=0$$
Using $J'_0=J_1$ and the same for $Y'_0$ we get a transcendental equation for eigenvalues $k_n$
$${Y_0(k_nr_1)\over J_0(k_nr_1)}={Y_1(k_nr_0)\over J_1(k_nr_0)}.\eqno(2.23)$$
The corresponding eigenfunctions are equal to
$$y_n(r)=\sqrt{r}[Y_0(k_nr_1)J_0(k_nr)-J_0(k_nr_1)Y_0(k_nr)]\equiv\sqrt{r}Z_0(k_nr).\eqno(2.24)$$
By general theorems about self-adjoint eigenvalue problems [3] the eigenfunctions form a complete orthogonal system in the Hilbert space $L^2([r_0,r_1])$. The normalization integral is easily calculated ([2], p.485)
$$\int\limits_{r_0}^{r_1}y_n^2(r)dr={1\over 2}[r_1^2Z_1^2(k_nr_1)-r_0^2Z_0²(k_nr_0)].$$
We denote the normalized eigenfunctions by $\fii_n(r)$. The solutions $k_n$ of (2.23) are approximately given by
$$k_n\approx {\al_n\over r_1},\eqno(2.25)$$
where $\al_n$,$n=0,1,2\ldots$ are the zeros of the Bessel function $J_0(z)$. This follows from the fact that the left-hand side of (2.23) assumes arbitrary values in the neighborhood of $\al_n$, and then can be made equal to the right-hand side. By the oscillation theorem [4] $\fii_n(r)$ has $n$ zeros in the interval $[r_0,r_1]$.

Now we are ready to write down the general solution of our ablation problem. The time dependent factor $T_3(t)$ follows from (2.16):
$\sim\exp(\omega_n^\pm t)$ where $\omega_n^\pm$ are the solutions of the quadratic equation
$$a\tau\omega^2+{\omega\over a}+k^2=0$$
namely
$$\omega_n^\pm=-{1\over 2\tau}\pm\sqrt{{1\over 4\tau^2}-{ak_n^2\over\tau}}.\eqno(2.26)$$
Then the total solution is equal to
$$T(t,r)=T_0+b\log{r_0\over r_1}\log{r\over r_1}-{b\over 2r^2}\log^2{r\over r_1}+$$
$$+{1\over\sqrt{r}}\sum_{n,\pm}a_n^\pm\fii_n(r)\exp(\omega_n^\pm t).\eqno(2.27)$$
The unknown coefficients $a_n^\pm$ are determined by the two initial conditions
$$T(0,r)=T_0,\quad {\d T\over\d t}(0,r)=f(r).\eqno(2.28)$$
In previous studies [1] the simple initial condition $f(r)=0$ has been assumed without any justification. However, since the heat generation $Q(r)$ (2.5) is $r$-dependent we expect $f(r)\ne 0$. Our experiments clearly show this (see next section). Using the two conditions (2.28) in (2.27) we obtain the two equations
$$\sum_n(a_n^++a_n^-)\fii(r)=-\sqrt{r}\B(b\log{r_0\over r_1}\log{r\over r_1}-{b\over 2r^2}\log^2{r\over r_1}\B)\eqno(2.29)$$
$$\sum_n(\omega_n^+a_n^++\omega_n^-a_n^-)\fii(r)=\sqrt{r}f(r).\eqno(2.30)$$
Since the eigenfunctions $\fii(r)$ form a complete orthonormal system the functions on the right side of (2.29) (2.30) can be uniquely expanded in the usual way by calculating scalar products, for example
$$\omega_n^1a_n^++\omega_n^-a_n^-=\int\limits_{r_0}^{r_1}\sqrt{r}f(r)\fii_n(r)dr$$
and similarly for (2.29). This finally allows to determine the coefficients $a_n^+$ and $a_n^-$ separately.

The form (2.27) of the solution is usefull to study the behavior for large and intermediate times $t$. Since the lowest eigenvalue $k_0$ is positive, $\omega_0^+$ in (2.29) is negative. Consequently, for $t\to\infty$ the solution goes exponentially towards the steady state. In the application to ablation in medicine one is interested in small $t$, as we will see in the next section. The analytic solution above may also be useful for testing numerical codes. But for small $t$ a different treatment must be used.

\section{Experimental study of the axisymmetrical ablation problem}

We have taken a stainless steel pot with radius $r_1=11$ cm with an isolating plate at the bottom. At the center $r=0$ we have placed a bar electrode of length $L=8.5$ cm and small radius $r_0=0.45$ cm. Filling the pot with water plus $0.3\%$ NaCl and applying a 13 Volt AC voltage of 50 Hz we have obtained a fairly accurate axisymmetric electric field as assumed in the last section. We measure the temperature in steps of 0.1 degree Celsius at various places $r$ as a function of time.

With help of a voltmeter we have first checked the radial dependence (2.3) of the AC potential. We found exact agreement with (2.3) for $r$ between 4 and 5 cm. For smaller and larger $r$ there are deviations of unknown origin. So we have restricted the temperature measurements to $r=4$ cm and 5 cm.

For each $r$ we have fitted the rise of the temperature $T(t)$ by means of a quadratic polynomial
$$T(t)-T_0=a_1t+a_2t^2.\eqno(3.1)$$
Let us try to describe the measured $T(t,r)$ by the usual first order heat equation
$$\d_tT={1\over\ro c}(\kappa\triangle T+Q).\eqno(3.2)$$
In this case we have at $t=0$
$$a_1=(\d_tT)(0)={Q(r)\over\ro c}.\eqno(3.3)$$
Calculating $Q$ from equation (2.5) there is a serious problem. The applied electric power $P_0$ which can be measured is only partly used for heating. Most of it goes into the electrolysis of Na Cl. In addition there appear bubbles of gas which also costs energy or power. In electrochemistry the heating power $P$ in (2.5) is written as
$$P=V_0\times I\eqno(3.4)$$
where $V_0$ is called overpotential and $I$ is the total current [5]. Since $V_0$ cannot be directly measured we treat it as a parameter to be determined.

Differentiating (3.2) we get for $t=0$
$$\d_t^2T={\kappa\over(\ro c)^2}\triangle Q(r)={\kappa\over(\ro c)^2}{1\over r}\d_r(r\d_r Q)=$$
$$={\kappa\over(\ro c)^2}{1\over r}\d_r\B(-{2\beta\over r^2}\B)={4\kappa\beta\over (\ro c)^2r^4}\eqno(3.5)$$
where (2.5) has been inserted. Since this is equal to $2a_2$ we can eliminate the unknown $\beta$
$${a_2\over a_1}={2\kappa\over\ro c r^2}.\eqno(3.6)$$
This relation is strongly violated by the measurements which shows that the simple heat equation (3.2) cannot be used for small times.

We proceed similarly with the hyperbolic heat equation
$$\d_t^2T=-{1\over\tau}\d_tT+{\kappa\over\tau\ro c}\triangle T+{Q\over\tau\ro c}.\eqno(3.7)$$
For $t=0$ we have
$$(\d_t^2T)(0)=-{a_1\over\tau}+{Q\over\tau\ro c}$$
or
$$2\tau a_2(r)-{Q(r)\over\ro c}=-a_1(r).\eqno(3.8)$$
The heat generation $Q(r)$ is given by (2.5)
$${Q\over\ro c}={P\over 4.185{\rm J}}{1\over 2\pi Lr^2}\B[{{\rm degrees}\over{\rm seconds}}\B]\eqno(3.9)$$
where $L$ and $r$ are measured in cm and $P$ in watts (4.185 J = 1 cal). The relevant power $P=V_0I$ (3.4) cannot be measured, therefore, we write equ.
(3.8) as
$$2\tau a_2(r)-{y\over r^2}=-a_1(r)\eqno(3.10)$$
with
$$y={60\,P\over 2\pi L\,4.185}\eqno(3.12)$$
where $P$ is the unknown heating power in watts. The factor 60 is inserted because $a_1$ in (3.10) is measured in degree Celsius/min. Now if we calculate $a_1(r)$ and $a_2(r)$ for two radial distances $r=\ro_1$ and $\ro_2$ from the temperature measurements, then we can determine $\tau$ and $y$.

The determination of $a_1$ and $a_2$ by a least square fit of the polynomial (3.1) to measured temperature values $T(t_j)$ is uncertain, depending on the number $N$ of values $t_j$ taken with. With our simple equipment (not very good thermometer) we have found for $N=4$
$$\ro_1=4 {\rm cm}:\quad a_1=0.122,\quad a_2=-5.86\cdot 10^{-3}$$
$$\ro_2=5 {\rm cm}:\quad a_1=0.0353,\quad a_2=1.59\cdot 10^{-3}$$
for $N=5$:
$$\ro_1=4 {\rm cm}:\quad a_1=0.116,\quad a_2=-3.36\cdot 10^{-3}$$
$$\ro_2=5 {\rm cm}:\quad a_1=0.0368,\quad a_2=1.36\cdot 10^{-3}$$
and for $N=6$:
$$\ro_1=4 \,{\rm cm}:\quad a_1=0.111,\quad a_2=-2.23\cdot 10^{-3}$$
$$\ro_2=5 \,{\rm cm}:\quad a_1=0.0386,\quad a_2=1.06\cdot 10^{-3}.$$
We see that $r=4$ cm is a critical distance where $a_2$ changes sign and more accurate measurements are necessary.

From the above results, equation (3.10) taken with $r=\ro_1$ and $\ro_2$ gives the following values for the interesting quantities $\tau$ and $y$:
$${\rm for}\,N=4:\quad\tau=4.0 \,{\rm min}\quad y=1.20$$
$${\rm for}\,N=5:\quad\tau=10.7 \,{\rm min}\quad y=2.57$$
$${\rm for}\,N=6:\quad\tau=6.5 \,{\rm min}\quad y=1.31$$
We see that the results obtained are quite uncertain due to the low quality thermometers which did not have the required sensitivity of 0.1 degree. The most reliable values are those of $N=6$ because they contain most measurements. From (3.12) we get $y=0.269 P$. Then the electric heating power is $P=y/0.269$ watts. For $y=1.31$ this gives 4.9 watts compared to 13 watts total applied power. This shows that most energy goes into the electrolysis and the bubble generation of gas at the central electrode.

For the medical application the value of the thermal relaxation time $\tau=6.5$ min is most important. In radiofrequency ablation one works with much smaller times $t=0.3-0.4$ min. On the one hand this makes life complicated because the hyperbolic heat equation must certainly be used. On the other hand this equation can simply be solved by a power series (3.1) in $t$. As we have seen the first two terms only depend on the heating power $Q(r)$. The heat conduction contributes to the cubic term
$$\d_t^3T=-{1\over\tau}\d_t^2T+{\kappa\over\tau\ro c}\triangle\d_t T.$$
If one has determined an approximate value of
$$(\d_tT)(0)=a_1(r)$$
one is able to calculate a correction
$$\triangle a
_1(r)={1\over r}\d_r(r\d_r a_1)$$
for heat conduction.

\end{document}